\documentclass[a4paper]{article}

\usepackage{INTERSPEECH2020}
\usepackage{url}
\usepackage{subfig}
\usepackage{balance}
\usepackage{amsmath}

\title{Should we hard-code the recurrence concept or learn it instead ? \\Exploring the Transformer architecture for Audio-Visual Speech Recognition}
\name{George Sterpu$^1$, Christian Saam$^2$, Naomi Harte$^1$}
\address{
  $^1$Sigmedia Lab, ADAPT Centre, School of Engineering, Trinity College Dublin, Ireland\\
  $^2$ADAPT Centre, School of Computer Science and Statistics, Trinity College Dublin, Ireland}
\email{\{sterpug, saamc, nharte\}@tcd.ie}

\begin{document}

\maketitle
\begin{abstract}
  The audio-visual speech fusion strategy AV Align  has shown significant performance improvements in audio-visual speech recognition (AVSR) on the challenging LRS2 dataset. Performance improvements range between 7\% and 30\% depending on the noise level when leveraging the visual modality of speech in addition to the auditory one. This work presents a variant of AV Align where the recurrent Long Short-term Memory (LSTM) computation block is replaced by the more recently proposed Transformer block. We compare the two methods, discussing in greater detail their strengths and weaknesses. We find that Transformers also learn cross-modal monotonic alignments, but suffer from the same visual convergence problems as the LSTM model, calling for a deeper investigation into the dominant modality problem in machine learning.
\end{abstract}

\section{Introduction}

Multimodal fusion ~\cite{baltrusaitis2019} allows the exploitation of redundancies and complementarities in naturally occurring signals, boosting the overall robustness of data processing systems in the presence of noise. One notable application is Audio-Visual Speech Recognition (AVSR), where the structure of the same speech signal materialises under two coherent modalities conveying variable levels of information. Recent advancements in machine learning led to a renewal of interest in AVSR and its key concepts such as cross-modal alignment and fusion~\cite{chung_cvpr_2017, Sterpu_ICMI2018, sterpu2020taslp}.

More recently, the AV Align strategy~\cite{Sterpu_ICMI2018, sterpu2020taslp}, which explicitly models the alignment of the audio and video sequences based on dot-product attention, has shown significant performance improvements on the challenging conditions of the LRS2 dataset~\cite{lrs2}. At the same time, AV Align also discovers block-wise monotonic alignments between the input modalities without alignment supervision, and outperforms a related strategy where the text modality is used as a proxy for fusion~\cite{chung_cvpr_2017}. This concept has also been applied to emotion recognition~\cite{tsai-etal-2019-multimodal, 9053762}, or speech grounding from video~\cite{paraskevopoulos2020}.

In the neural network space, input sequences are traditionally processed by recurrent neural networks variants including Long Short-term Memory networks (LSTM~\cite{lstm}). A more recent model called Transformer~\cite{transformers} removes the recurrent connections and updates the sequence representations using instead self-attention connections. Without the sequential processing constraint of RNNs, the Transformer model can better leverage computational resources through parallelism and achieve performance comparable to LSTMs on speech recogntion tasks for a fraction of the training costs~\cite{zeyer_asru2019,9003750, 8462506}.

In this work, we explore a variant of the AV Align strategy where the LSTM cells previously used in ~\cite{Sterpu_ICMI2018, sterpu2020taslp} are replaced with Transformer layers. Our contributions are as follows. We describe the equivalent Transformer block for the alignment operation of AV Align in Section~\ref{sec:method}. We train and evaluate Audio-only and Audio-Visual Transformer models on the LRS2 dataset under identical conditions with~\cite{sterpu2020taslp} in Section~\ref{sec:avperf}. In Section~\ref{sec:alignments} we show that the Audio-Visual Transformer suffers from the same video convergence problem as the LSTM-based AV Align, and the auxiliary Action Unit loss helps recover the previously seen performance improvements.


\begin{figure}
    \centering
    \includegraphics[width=0.9\linewidth]{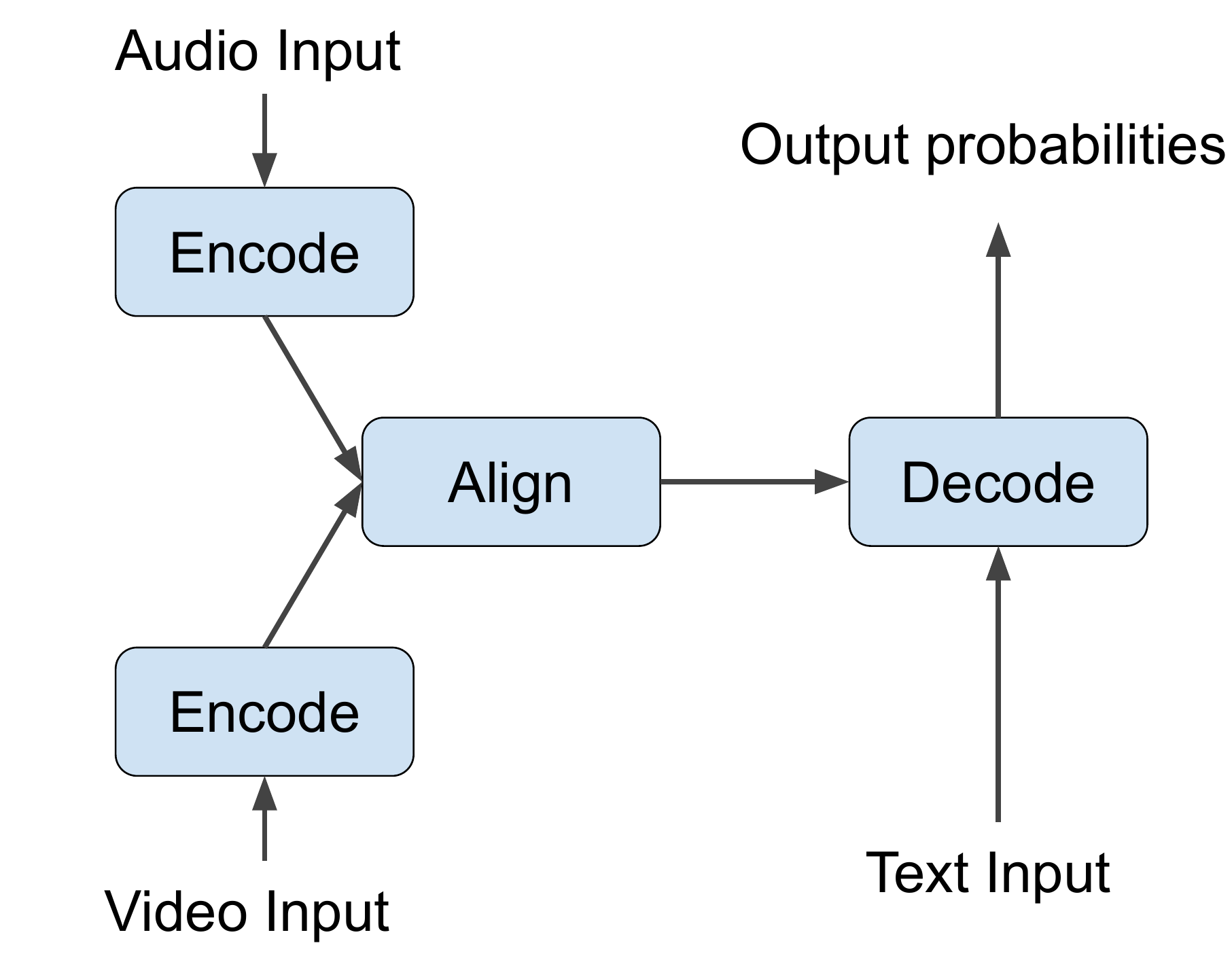}
    \caption{Schematic representation of the Audio-Visual alignment and fusion strategy AV Align for sequence to sequence speech recognition.}
    \label{fig:avalign}
\end{figure}

\section{Audio-Visual Transformer}
\label{sec:method}

The AV Align architecture was first introduced in \cite{Sterpu_ICMI2018} and treated in more detail in \cite{sterpu2020taslp}. The aim of this section is to show how the original LSTM block in AV Align can be replaced with the Transformer one.

In the most general case, given a variable length acoustic sentence $a = \{a_1, a_2, \ldots , a_N\}$ and its corresponding visual track $v = \{v_1, v_2, \ldots , v_M\}$, we transform the raw input signals into higher level latent representations (denoted by ${o}_{A} = \{o_{A_1}, o_{A_2}, \ldots , o_{A_N}\}$ and ${o}_{V} = \{o_{V_1}, o_{V_2}, \ldots , o_{V_M}\}$) using stacks of Transformer Encoders. An Align stack introduced in Section~\ref{sec:alignstack} soft-aligns the two modalities, and computes the fused sequence $o_{AV}$. Finally, an auto-regressive decoder is used to predict the output sequence of graphemes. 

The Transformer architecture defined in \cite{transformers} is made of an Encoder and a Decoder stack. The Encoder stack contains repeated blocks of self-attention and feed forward layers. The decoder stack contains repeated blocks of self-attention, decoder-encoder attention, and feed-forward layers. The inputs to these stacks are summed with positional encodings to embed information about the absolute position of timesteps within sequences. The Encoder and the Decoder stacks are schematically illustrated in Figure~\ref{fig:stacks}.

\subsection{The Align stack}
\label{sec:alignstack}

\begin{figure}
    \centering
    \includegraphics[width=\linewidth]{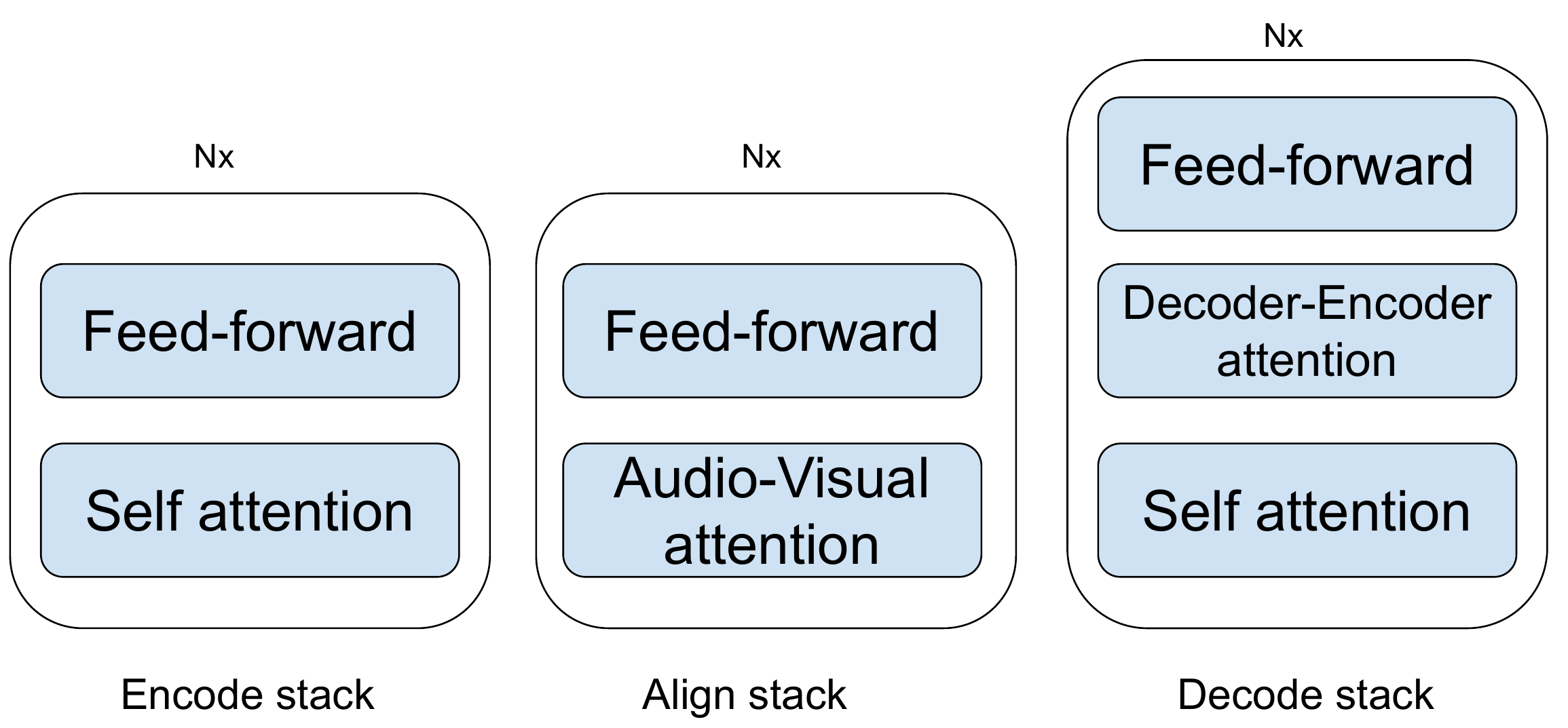}
    \caption{The three main blocks of the Audio-Visual Transformer variant. The Encode and Decode stacks are the same as in the original model~\cite{transformers}. We introduce the Align stack which is based on generic attention and feed-forward layers. We keep the Nx notation from the original article to imply stacking together multiple blocks of the same structure.}
    \label{fig:stacks}
\end{figure}

In order to adapt AV Align to the Transformer architecture, we define an additional Align stack as the repeated application of cross-modal attention and feed-forward layers. The Align stack is displayed in Figure~\ref{fig:stacks} between the Encode and the Decode stacks.
The cross-modal attention layer is a generic attention layer applied between the outputs of the two stream encoders. The align block takes video outputs $o_V$ and audio outputs $o_A$ as keys and queries respectively, whereas the regular encoder-decoder attention layer receives audio representations and graphemes. Consequently, the Audio-Visual Transformer model implements a single generic attention operation, as originally defined in~\cite{transformers}, maintaining simplicity.

Formally, the audio-visual alignment and fusion steps of the attention layer in the Align stack can be described as:
\begin{align}
    c_V  &= \mathsf{attention}(\mathrm{query}=o_A, \mathrm{source}=o_V) \label{eq:aln}\\
    o_{AV} &= c_V + o_A & \label{eq:fuse}
\end{align}
where $c_V$ are the visual context vectors computed as linear combinations of the video source $o_V$. 
Both the LSTM and Transformer variants of AV Align use the concept of dot-product attention to align the higher level audio and video representation, as in (\ref{eq:aln}). However, whereas the LSTM model~\cite{Sterpu_ICMI2018} fuses the visual context vector with audio representation by concatenating them and projecting to a shared space using linear combination, the fusion operation in the Transformer is a residual connection, seen in (\ref{eq:fuse}). Adding one layer's inputs to the attention output is the default fusion mechanism of the original Transformer model~\cite{transformers}, also being used to fuse the decoder's input queries with the audio/audiovisual keys. A linear combination style fusion was explored internally for the Transformer model, without significant findings. No statistically significant differences were also reported in~\cite{sterpu2020taslp} when exploring multiple feature fusion strategies.

 Our implementation forks the Transformer model officially supported in TensorFlow 2~\cite{tf_transformer} and only adds the high level Align stack, together with the visual convolution front-end, reusing the existing implementations of attention and feed-forward layers. Our code will be made publicly available\footnote{https://github.com/georgesterpu}.

Compared to the multi-modal Transformer model proposed in~\cite{tsai-etal-2019-multimodal}, we do not make use of cross-modal attention at every layer in the alignment stack. As we argued in~\cite{Sterpu_ICMI2018}, there may be limited correspondences between audio and video at the lower levels of representations, and aligning only the higher level concepts is likely to speed up the training convergence.  
\section{Experiments and Results}

\subsection{Setup}

\textbf{LRS2} \cite{lrs2} contains 45,839 spoken sentences from BBC television recorded in uncontrolled illumination conditions, challenging head poses, and a low image resolution of 160x160 pixels. LRS2 is the largest AVSR dataset publicly available for research, and allows the comparison of results with more recent work~\cite{8639643, 8585066, 9054127}.

Our system takes auditory and visual input concurrently. The \textbf{audio} input is the raw waveform signal of an entire sentence. The \textbf{visual} stream consists of video frame sequences, centred on the speaker's face, which correspond to the audio track. We use the OpenFace toolkit \cite{openface2} to first  detect and align the faces. We then crop the lip area to a static window determined heuristically, covering the bottom 40\% of the image height and the middle 80\% of the image width.

\textbf{Audio input.} The audio waveforms sampled at 16,000 Hz. Following the procedure of~\cite{sterpu2020taslp}, we add cafeteria acoustic noise to the clean signal at three different Signal to Noise Ratios (SNR) of 10db, 0db, and -5db, in order to study how the audio information loss affects learning. We compute the log magnitude spectrogram of the input, choosing a frame length of 25ms with 10ms stride and 1024 frequency bins for the Short-time Fourier Transform (STFT), and a frequency range from 80Hz to 11,025Hz with 30 bins for the mel scale warp. We stack the features of 8 consecutive STFT frames into a larger window, leading to an audio feature vector $a_i$ of size 240, and we shift this window right by 3 frames, thus attaining an overlap of 5 frames between windows.

\textbf{Visual input.} We down-sample the 3-channel RGB images of the lip regions to 36x36 pixels. A ResNet CNN \cite{resnet2} processes the images to produce a feature vector $v_j$ of \textbf{256 units} per frame. The network architecture is the same as in ~\cite{sterpu2020taslp}, and is detailed in Table~\ref{tab:resnet_details}.

\begin{table}[t]
\centering
\caption{CNN Architecture. All convolutions use 3x3 kernels, except the final one. The Residual Block is taken from \cite{resnet2} in its \emph{full preactivation} variant.}
\label{tab:resnet_details}
\begin{tabular}{rcr}
\textbf{layer} & \textbf{operation}                                      & \textbf{output shape} \\ \hline
0              & Rescale [-1 ... +1]                                                & 36x36x\textbf{3}               \\ 
1              & Conv                                                    & 36x36x\textbf{8}               \\ 
2-3            & Res block                                               & 36x36x\textbf{8}               \\ 
4-5            & Res block                                               & 18x18x\textbf{16}              \\ 
6-7            & Res block                                               & 9x9x\textbf{32}                \\ 
8-9            & Res block                                               & 5x5x\textbf{64}                \\ 
10             & Conv 5x5                                                & 1x1x\textbf{128}
\end{tabular}
\end{table}

\subsection{Neural network details}

The transformer model uses 6 layers in the Encoder and Decoder stacks, a model size $d_{model} = 256$, a filter size $d_{ff} = 256$, one attention head, and 0.1 dropout on all attention weights and feedforward activations. The Align stack is made of a single block of cross-modal attention and feed-forward layers, with one attention head. We performed an ablation study, noting that an increase in width and depth was not worth the additional computation time with respect to accuracy.

\subsection{Audio-Visual Speech Recognition Performance}
\label{sec:avperf}

We train both audio and audio-visual Transformer models on the LRS2 dataset, with the audio modality corrupted in four stages of cafeteria noise. As in~\cite{sterpu2020taslp}, we train an additional audio-visual model with the Action Unit loss enabled. The results are shown in Table~\ref{tab:avt}.

\begin{table}[th]
  \caption{Character Error Rate [\%] on LRS2}
  \label{tab:avt}
  \centering
  \begin{tabular}{ l l l l r }
    \toprule
    \multicolumn{1}{c}{\textbf{System}} & \multicolumn{1}{c}{\textbf{clean}} &
    \multicolumn{1}{c}{\textbf{10db}} &
    \multicolumn{1}{c}{\textbf{0db}} &
    \multicolumn{1}{c}{\textbf{-5db}}\\
    \midrule
    Audio LSTM~\cite{sterpu2020taslp} & 16.38 & 21.85 & 36.27 & 49.08 \\
    AV Align LSTM + AU~\cite{sterpu2020taslp} & 15.57 &	18.28 &	26.57 &	33.98 \\
    \midrule
    Audio Transformer & 13.65 & 18.84 & 31.21 & 43.71 \\
    AV Transformer &  13.06 & 17.77 & 30.60 &  43.28\\
    AV Transformer + AU & 12.08 & 14.82 & 23.52 & 31.74 \\
    \bottomrule
  \end{tabular}
  
\end{table}

We notice that the AV Transformer achieves a similar performance to the Audio Transformer, suggesting that the video modality was uninformative. We start seeing performance improvements only when the AU loss is used, reproducing the finding in ~\cite{sterpu2020taslp}. 
The relative performance improvements of the AV Transformer + AU over the Audio Transformer start at 7.5\% in clean speech, and go up to 26.6\% in noised speech. Thus, the visual modality brings similar levels of relative improvements over the audio-only modality to both the Transformer and the LSTM trained in~\cite{sterpu2020taslp}. The absolute error differences between the LSTM and the Transformer models are partly owed to the larger model size of the Transformer used in this work (25 MB Audio, 36 MB AV) over the LSTM one in~\cite{sterpu2020taslp} (9.3 MB Audio).

\subsection{Audio-Visual Alignments}
\label{sec:alignments}

\begin{figure}[t]
        \centering
        \subfloat[Base model]{
           \includegraphics[width=0.4\linewidth]{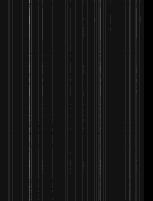}
           \label{fig:aln:noau}
        }
        \subfloat[Base model + AU Loss]{
        \includegraphics[width=0.4\linewidth]{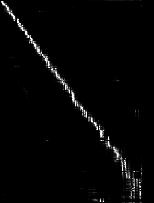}
           \label{fig:aln:au}
        }
        \caption{The Audio-Visual alignments learnt by the Transformer models}
        \vspace{-0.1337cm}
    \label{fig:alignments}
    \end{figure}

As in~\cite{sterpu2020taslp}, we inspect the alignment weights between the audio and visual representations, which are displayed in Figure~\ref{fig:alignments}. Without the AU Loss, the AV Transformer has the same difficulty as the Audio-Visual LSTM of~\cite{sterpu2020taslp} to learn cross-modal correspondences (Figure~\ref{fig:aln:noau}), thought to be caused by the improper learning of visual representations. The alignments emerge as monotonic at the macro-block level with the AU loss (Figure~\ref{fig:aln:au}).

\section{Transformer or LSTM for AVSR?}

The results show that the self-attention connections of the Transformer model can successfully substitute the recurrent ones originally used in the LSTM-based AV Align~\cite{Sterpu_ICMI2018}. As in~\cite{sterpu2020taslp}, the cross-modal alignments emerge as locally monotonic based on the dot-product correlations between audio and video representations. Without the auxiliary Action Unit loss, the AV Transformer presents the same learning difficulties as the LSTM variant of AV Align, and does not manage to learn monotonic alignments. We have previously speculated that the convergence problem of the visual module in AV Align was partly due to the longer propagation path of the error signal for the visual CNN and RNN in the sequence to sequence structure. Despite the great reduction of this path length in a Transformer network, our AV Transformer still required the AU loss. This demands a deeper investigation into the dominant modality problem in multi-modal machine learning, where patterns need to be discovered in the weaker visual signal. Also seen in~\cite{Sterpu_ICMI2018, sterpu2020taslp}, sequence to sequence models are known to be susceptible to encoder-decoder disconnect when the information distribution in the target signal can be exploited for localised optimisation of the decoder, and the audio-visual disconnect is another ramification of the same problem. 


Our study does not reflect an analysis of the parameter efficiency of the Transformer network compared to the LSTM for this particular dataset. We opted for commonly used hyper-parameters for datasets of this size, noting that the Transformer model is larger, partly explaining the improvements in error rates. This is because the advantages and disadvantages of both strategies go beyond parameter efficiency, being reflected in hardware throughput and engineering effort, and are discussed in greater detail in~\cite{zeyer_asru2019}.

It has been suggested before that Transformers do learn the concept of recurrence from self attention connections. However, despite their highly parallel design conveying significant performance advantages over LSTMs, there is still a sense of wastefulness, particularly in speech, where distant inputs are unlikely to require connectivity. Additionally, the information from one speech frame to another does not change so much as to demand a full update of every representation in a layer.
    
Despite these inefficiencies, the Transformer architecture achieves faster computation speeds than LSTM on modern hardware for the majority of today's benchmarks. The LSTM blocks are facing more technical and engineering challenges in modern machine learning frameworks, which additionally leads to higher maintenance and development costs. In~\cite{sutton2019bitter} it is argued that general purpose algorithms that best leverage computation scaling appear to be the most successful ones in the long run. The quintessential question becomes: is recurrence a concept that we want to embed into neural networks by hand, or is it preferable to opt for simpler architectures that allow the automatic learning of it ?

\section{Acknowledgements}
Our work is supported by a GPU grant from NVIDIA. The ADAPT Centre for Digital Content Technology is funded under the SFI Research Centres Programme (Grant 13/RC/2106) and is co-funded under the European Regional Development Fund.

\balance
\bibliographystyle{IEEEtran}
\bibliography{mybib}

\end{document}